\documentclass[conference]{IEEEtran}
\IEEEoverridecommandlockouts
\usepackage{cite}
\usepackage{amsmath,amssymb,amsfonts}
\usepackage{algorithmic}
\usepackage{graphicx}
\usepackage{hyperref}
\usepackage{subcaption}
\usepackage{caption}
\usepackage[font=footnotesize]{caption} % or scriptsize/small
\usepackage{textcomp}
\usepackage{xcolor}
\usepackage{booktabs}
\usepackage{array}
\usepackage[table]{xcolor} % for \cellcolor

\usepackage{colortbl}

\definecolor{EVgreenA}{RGB}{169,210,127}
\definecolor{EVgreenB}{RGB}{99,190,123}
\definecolor{EVgreenC}{RGB}{160,208,127}
\definecolor{EVyellowA}{RGB}{255,235,132}
\definecolor{EVsalmon}{RGB}{249,134,112}
\definecolor{EVyellowB}{RGB}{247,233,132}
\definecolor{EVorangeA}{RGB}{253,214,127}
\definecolor{EVredA}{RGB}{248,105,107}
\definecolor{EVorangeB}{RGB}{252,193,123}

\definecolor{EVyellow}{RGB}{255,235,132}     
\definecolor{EVgreen}{RGB}{99,190,123}       
\definecolor{EVsalmonA}{RGB}{249,133,112}    
\definecolor{EVsalmonB}{RGB}{248,119,109}    
\definecolor{EVygreen}{RGB}{216,224,130}     
\definecolor{EVred}{RGB}{248,105,107}

\def\BibTeX{{\rm B\kern-.05em{\sc i\kern-.025em b}\kern-.08em
    T\kern-.1667em\lower.7ex\hbox{E}\kern-.125emX}}
\begin{document}

\title{\huge Voltage and Frequency Stability Analysis of Transmission Power Grids with EV Charging Stations\\
}

\author{%
\small
\hspace*{-4.5em}%

\begin{tabular}{c}
\begin{tabular}{c} % Nested table for the names only
Akib Mostabe Refat$^{1}$,
Mohammed F. Al-Mashdali$^{1}$,
Alan Cordic$^{1}$,
Abdulaziz Qwbaiban$^{1}$,  
Emad Abukhousa$^{4}$, \\   % Break happens here
Kazi N. Hasan$^{5}$,
M. A. Abido$^{1,2,3}$,
Mohammed Al-Muhaini$^{1}$
\end{tabular}\\[0.6em] % Vertical space after the names block
$^{1}$Electrical Engineering Department, King Fahd University of Petroleum \& Minerals (KFUPM), Dhahran 31261, Saudi Arabia\\
$^{2}$Interdisciplinary Research Center for Sustainable Energy Systems (IRC-SES), KFUPM, Dhahran 31261, Saudi Arabia\\
$^{3}$SDAIA-KFUPM Joint Research Center for Artificial Intelligence, KFUPM, Dhahran 31261, Saudi Arabia\\
$^{4}$School of Electrical and Computer Engineering, Georgia Institute of Technology, Atlanta, GA, USA\\
$^{5}$School of Engineering, RMIT University, Melbourne 3001, Australia\\[0.4em]
\textit{\{g202315250, g201864560, g202525710, qwbaiban, mabido, muhaini\}@kfupm.edu.sa};\
\textit{emadak@gatech.edu};\ 
\textit{kazi.hasan@rmit.edu.au}
\end{tabular}
}

\maketitle

\begin{abstract}
The large-scale Electric Vehicle (EV) integration into the electricity grid has initiated significant challenges to grid stability issues due to dynamic loadability events. Although electric vehicle impacts on distribution systems are well studied, transmission-level investigations remain limited. In this research paper, case scenarios of EV load models as charging stations have been considered for stability analysis (Voltage and Frequency Stability) to address EV operation on the transmission grid. It is also noted that the operation of EV stations due to their high loadability causes more stability complexities to the grid compared to other loads in a power network. Simulations have been conducted on two different power networks of the IEEE-9 and IEEE-39 bus test systems, respectively.
\end{abstract}


\section{Introduction}
Electric Vehicle (EV) integration into transmission networks significantly affects grid stability, especially at vulnerable buses. To evaluate variation in power and system voltage, EVs alter grid stability and should be modeled as three-phase dynamic loads. Stability analysis is necessary to evaluate real and reactive power loading, weak lines and grid oscillations.

Recent research has explored the combined impact of EVs and renewable sources such as Photovoltaic (PV) systems, on power quality and stability [1]. Maintaining voltage within $±5\%$–$6\%$ of nominal value after disturbances [2][3] and a power factor of at least 0.95 are vital for secure grid operation [4]. Tavakoli et al. [1] analyzed the joint PV–EV impact and emphasized coordinated control to sustain grid stability. However, most studies focus on distribution-level power quality or steady-state voltage. Recent probabilistic and deep-learning methods improve voltage-stability prediction in renewable networks but are seldom applied to EV integrated transmission systems [5][6]. Momoh et al. [7] have articulated the future directions on Single Stage Converter (SSC) technique to address the State of Charge (SoC) problems of EV charging stations in maintaining grid stability.

This research focuses on the voltage and frequency stability of EV integrated transmission networks using IEEE-9-bus and IEEE-39-bus test systems. Three EV load models are developed to analyze their grid impact. Voltage stability is assessed using P–V curves to address voltage collapse under heavy loading, the Fast Voltage Stability Index (FVSI), and the Novel Line Stability Index (NLSI) to identify weak transmission lines and determine loadability limits. Frequency stability, primarily governed by generator inertia, is defined as maintaining synchronism by sustaining core frequency parameters following grid disturbances and power imbalances [2]. The North American Electric Reliability Corporation (NERC) report highlights key frequency-response concerns and stresses the need for simulation case studies for interconnection stability [8]. It is analyzed for the IEEE-9-bus system under grid disturbances and EV load placements using two main metrics—Frequency Nadir and Rate of Change of Frequency (ROCOF). Multi-Band PSS (MB-PSS) and Automatic Generation Control (AGC) are also examined to show improved damping and frequency regulation, guiding EV load placement in weak grids by explaining post-disturbance restoration.

To address rising EV demand, this study focuses on the stability limitations since very few research works have been conducted to consider EV charging stations in transmission grids [9]. This research contributes to the assessment of grid impacts on voltage and frequency stability to facilitate large-scale EV charging station integration, through the development of appropriate EV load models and the identification of optimal modeling parameters.

The remainder of this paper is organized as follows: Section II presents the modeling framework and methodology for EVs and transmission networks, respectively; Section III discusses simulation results and solutions for voltage and frequency stability; and Section IV concludes the paper with major findings and recommendations for future work.

\section{Methodology and modeling}
This paper concentrates on EV station modelling approaches for stability analysis of grid integration using the IEEE-9 and IEEE-39 power networks as test cases. MATLAB/Simulink software models the transmission network and dynamic EV loads have been developed.

\subsection{Systematic modeling of EV load for grid integration}\label{AA}
To realize the stability analysis problem, a simple 2-bus system scenario incorporating an EV and a load is considered. The load includes a constant-power type and EV, with power transfer between buses.

Previously, EV load was regarded as a Constant Impedance, Current, and Power (ZIP) model. Tavakoli et al. [1] defined the EV station as a dynamic load by including the constant power components $a$ and $b$ and a voltage negative exponential term $\alpha$, as shown in equation (1). The parameters described in (1) have been utilized in different EV models under dynamic three phase load block. The actual dynamic load equation in (2) is predefined in the SIMULINK workspace, where exponent, $n_p$ controls the load type and $T_p$, $T_s$ are the time constants representing active power in (2) [10].

\begin{equation}
\frac{P}{P_0}=a(\frac{V}{V_0})^{\alpha}+b\label{eq}
\end{equation}

\begin{equation}
\frac{P}{P_0}=(\frac{V}{V_0})^{n_p}(\frac{1+T_p}{1+T_s})\label{eq}
\end{equation}

where $P$,$V$,$P_0$, and $V_0$ indicate real power consumption and voltages with existing and initial measurements on the EV load bus, respectively. Therefore, the following EV load model constraints (3) - (6) are described in detail to ensure the system stability of the power network model. It is to be noted that the value of power components and lead resistance, $R_L$ varies based on the power network type.
\begin{align}
\alpha &\leq -1 \\
0.05 &\leq a \leq 0.07 \\
0.93 &\leq b \leq 0.95 \\
0.01\,\Omega &\leq R_L \leq 0.09\,\Omega
\end{align}

By considering the number and system constraints of dynamic EV model for grid integration in such a way to ensure the following power system constraints (7) - (9).
\begin{gather}
\text{\textit{FVSI,NLSI}} < 1 \\
59.5\,\text{Hz} \leq f_{\text{nom}} \leq 60.1\,\text{Hz} \\
\text{\textit{ROCOF}} < 1\,\text{Hz/s}
\end{gather}

\subsection{Description of the power transmission network model}\label{AA}
This section presents a summary of test bus systems on large-scale EV load model integration affecting grid stability. Stability studies are performed on the IEEE-9 and IEEE-39 bus networks described in [5][11], with Table - 1 listing key system parameters for load-flow analysis. The P–V curve algorithm from [6] is implemented in this research.

\begin{table}[htbp]
\centering
\caption{\textbf{Transmission Network System Parameters}}
\label{tab:wide_table}
\begin{tabular}{|c|c|}
\hline
\textbf{Parameter} & \textbf{Value} \\
\hline
Frequency \textit{fsys} & 60 Hz \\
\hline
Base Voltage \textit{Vb} (IEEE-9) & 230 KV \\
\hline
Base Voltage \textit{Vb} (IEEE-39) & 345 KV \\
\hline
Solar Irradiance \textit{$G_{rated}$} & $1000 \ \text{W}/\text{m}^2$ \\
\hline
\end{tabular}
\end{table}

To identify weak transmission lines prone to congestion under EV load integration, suitable stability indices are applied. Following Salama and Vokony [12], the FVSI and NLSI in (10) and (11) are adopted, and their values must remain below unity to avoid system instability. The equations with subscripts \textit{i} and \textit{j} represent the sending and receiving end busses, respectively.

\begin{equation}
FVSI=\frac{4Z_{ij}^{2}Q_j}{V_j^{2}X_{ij}}\label{eq}
\end{equation}

\begin{equation}
NLSI=\frac{R_{ij}P_j+X_{ij}Q_j}{0.25V_j^{2}}\label{eq}
\end{equation}

\subsection{EV load modeling approaches in transmission network planning}
EV load integration into the power grid significantly affects stability, especially at weak buses. This section examines three EV models under varying loading conditions. Grid stability depends on load variation and type. Three 5 kV EV models, each linked to a 5 km transmission line, are used for several case studies. Each model is combined with a PV farm of 1 km line to enhance stability and analyzed using the following two equations (12) and (13) in [13].

\begin{equation}
P_{PV}=P_b * \frac{G_{Actual}}{G_{Rated}}\label{eq}
\end{equation}

\[
G_{\text{Actual}} =
\begin{cases}
600 + \text{rand}\,(G_{\text{rated}} - 500); & \text{if } \text{rand} > 0.6 \\
0; & \text{rand} < 0.6
\end{cases}
\]

\begin{equation}
Q_{PV}=0.2*P_{PV}\label{eq}
\end{equation}

where, $P_{PV}$ and $Q_{PV}$ represents the active and reactive power module of PV farm. The actual and rated irradiance of $1000\frac{W}{{m}^{2}}$ using data randomization is unpredictable depicting uncertain generation due to varying weather conditions.

The first EV model with lead resistance is connected to the charging infrastructure, as shown in Fig. 1. The setup includes charging limits, energy calculations, and reactive power capability based on rated EV power [14]. The SoC limits are set to $90\%$ and $10\%$, with overall efficiency of $96\%$. The reactive power capability module, linked to external PQ control, provides saturated reactive power by comparing calculated and base real power of the charging module. In the Stored Energy Calculation block, SoC (\%) is obtained from EV energy integration based on charging-block power, considering both charging and discharging efficiencies.

\begin{figure}[htbp]
\centerline{\includegraphics[width=3.6in,height=1.8in]{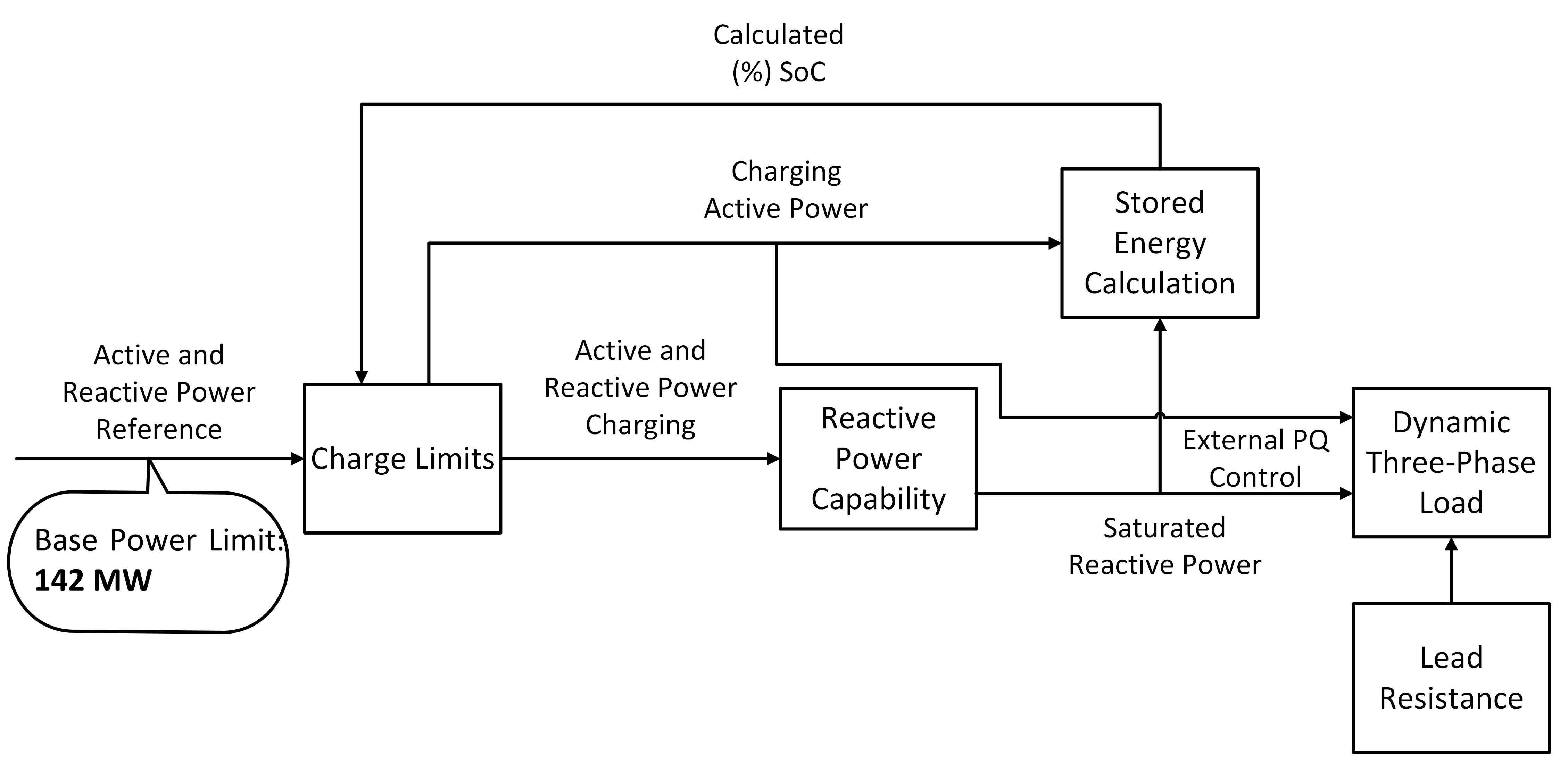}}
\caption{Block Diagram of First EV Model with Charging Module.}
\label{fig}
\end{figure}

In the second EV model at a particular lead resistance, as shown in Fig. 2, the dynamic load power control is applied to the external PQ control of the EV load with a power factor of 0.95 to obtain reactive power by varying the initial real power. In the dynamic load controller block, a step function varies the real and reactive power according to the rated real power of the EV load [15].

\begin{figure}[htbp]
\centerline{\includegraphics[width=2.6in,height=1.3in]{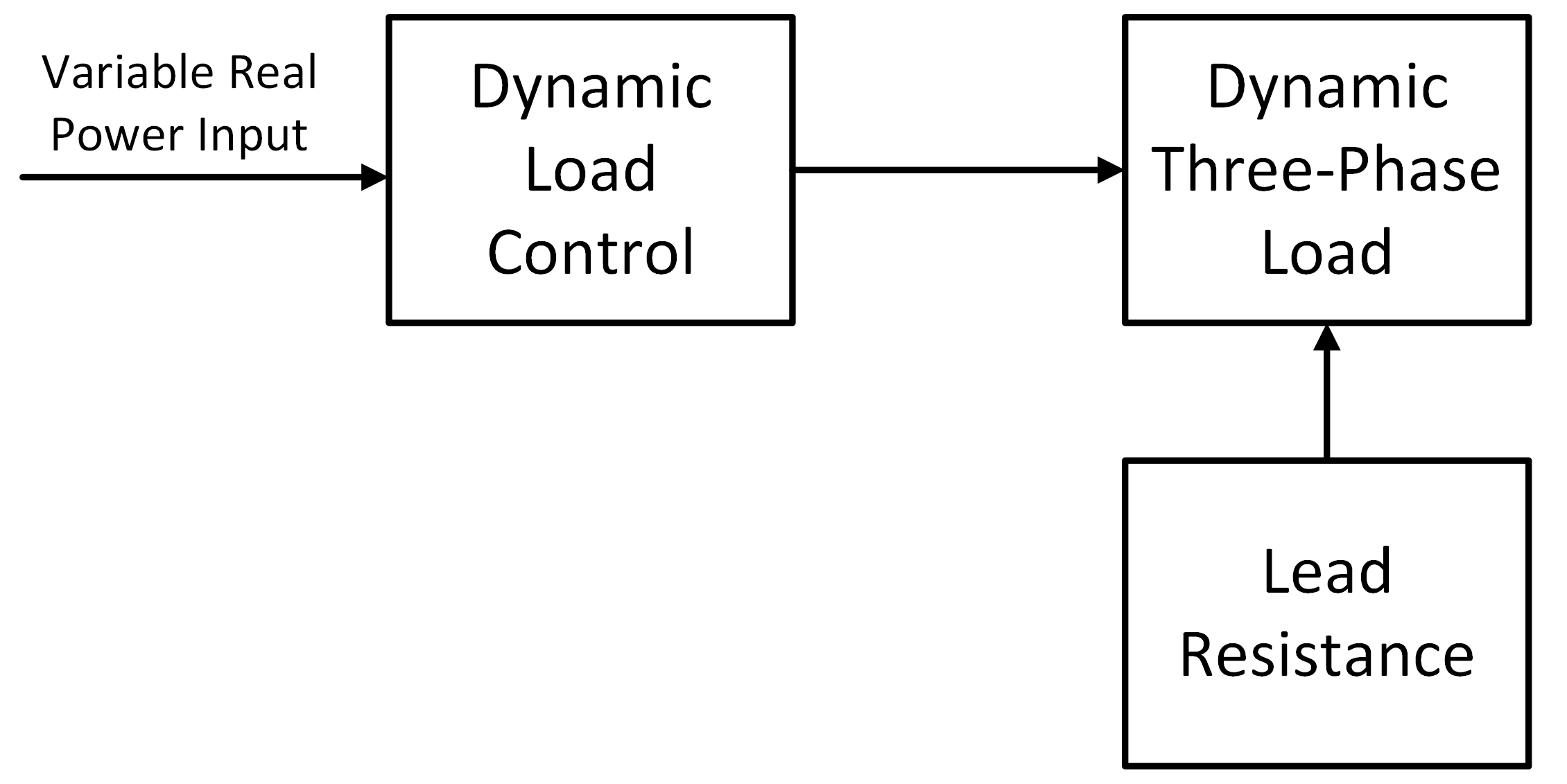}}
\caption{Second EV Model with Dynamic Load Control Module.}
\label{fig}
\end{figure}

However, no such control method has been used by analyzing the third model. This model has been prescribed to visualize the changes by maintaining the power and voltage exponential component owing to the load variation of the EV model [1][16].

\subsection{Modeling of synchronous generator and power system stabilizer}
For frequency stability assessment, synchronous generators with AGC and frequency regulation are modeled at the slack and PV buses to study the grid impact of EV models. The synchronous machine, with its mechanical and electrical control systems, converts mechanical power $P_m$ from the Steam Turbine and Governor (STG) into three-phase electrical power, while the field voltage ${V_f}$ regulates internal emf for real-power and frequency control.

The steam turbine and voltage regulator supply $P_m$ and $V_f$ to the generator based on reference power and voltage in p.u., $P_{ref}$ and $V_{ref}$ with the PSS status. The AGC maintains steady-state frequency at 60 Hz after a grid disturbance [17], monitoring rotor-speed deviation, $d\omega$, and computing power deviation $\Delta P$ for frequency regulation. The frequency block calculates nominal frequency $f_{nom}$, nadir $f_{nadir}$ and ROCOF ,$\frac{df}{dt}$, from $d\omega$ and $\Delta P$. Together, the generating system coordinates AGC, voltage excitation, and PSS to regulate grid frequency.

The PSS enhances damping and improves grid frequency after disturbances. The IEEE type multi-band PSS4B derives its stabilizing signal from rotor-speed deviation $d\omega$ and feeds it to the voltage regulator, as implemented in Simscape and described in MATLAB documentation [18]. The frequency response magnitude is tuned using gain blocks for each band (LF = 0.2 Hz, IF = 1.25 Hz, HF = 12 Hz).

The transfer function for MB PSS based on conceptual representation with washout and lead–lag compensation is derived using (14) - (17) [19]. The PSS4B improves damping by combining contributions from different frequency ranges while filtering DC offsets and non-oscillatory components.

\begin{equation}
u_{PSS} = \left[ G_{B} + G_{I} + G_{H} \right] \Delta \omega
\label{eq:upss}
\end{equation}

\begin{equation}
G_{B} = K_{B}\,\frac{sT_{BW}}{1+sT_{BW}}\,
        \frac{1+sT_{B1}}{1+sT_{B2}}\,
        \frac{1+sT_{B3}}{1+sT_{B4}}
\label{eq:GB}
\end{equation}

\begin{equation}
G_{I} = K_{I}\,\frac{sT_{IW}}{1+sT_{IW}}\,
        \frac{1+sT_{I1}}{1+sT_{I2}}\,
        \frac{1+sT_{I3}}{1+sT_{I4}}
\label{eq:GI}
\end{equation}

\begin{equation}
G_{H} = K_{H}\,\frac{sT_{HW}}{1+sT_{HW}}\,
        \frac{1+sT_{H1}}{1+sT_{H2}}\,
        \frac{1+sT_{H3}}{1+sT_{H4}}
\label{eq:GH}
\end{equation}

\section{Simulation results and proposed solutions}

In this section, the stability assessment for the IEEE 9 and 39-bus system evaluates EV load model to grid scenarios. The maximum number of EV stations are connected to the buses based on voltage profile analysis of critical buses of the power networks. In this case study, EV models are placed at buses 7, 8 and 9 for IEEE 9-bus system and at buses 15, 17, 26 and 27 for IEEE 39-bus system respectively.

\subsection{Voltage stability findings}
This research analyzes the grid impact of EV models on test bus systems at weaker buses in transmission networks. Voltage profile analysis identifies optimal EV load placement, while stability results reveal weak grid sections. The following Fig. 3 indicates the lower bus voltages of IEEE 9 and 39 bus system at critical power point loadability. For each model, P–V curves and indices locate the critical power point and lines prone to collapse. Comparative results show that coordinated control and reactive power support of EV models shift the critical point, improving system stability.

\begin{figure}[htbp]
\centerline{\includegraphics[width=3.6in,height=1.7in]{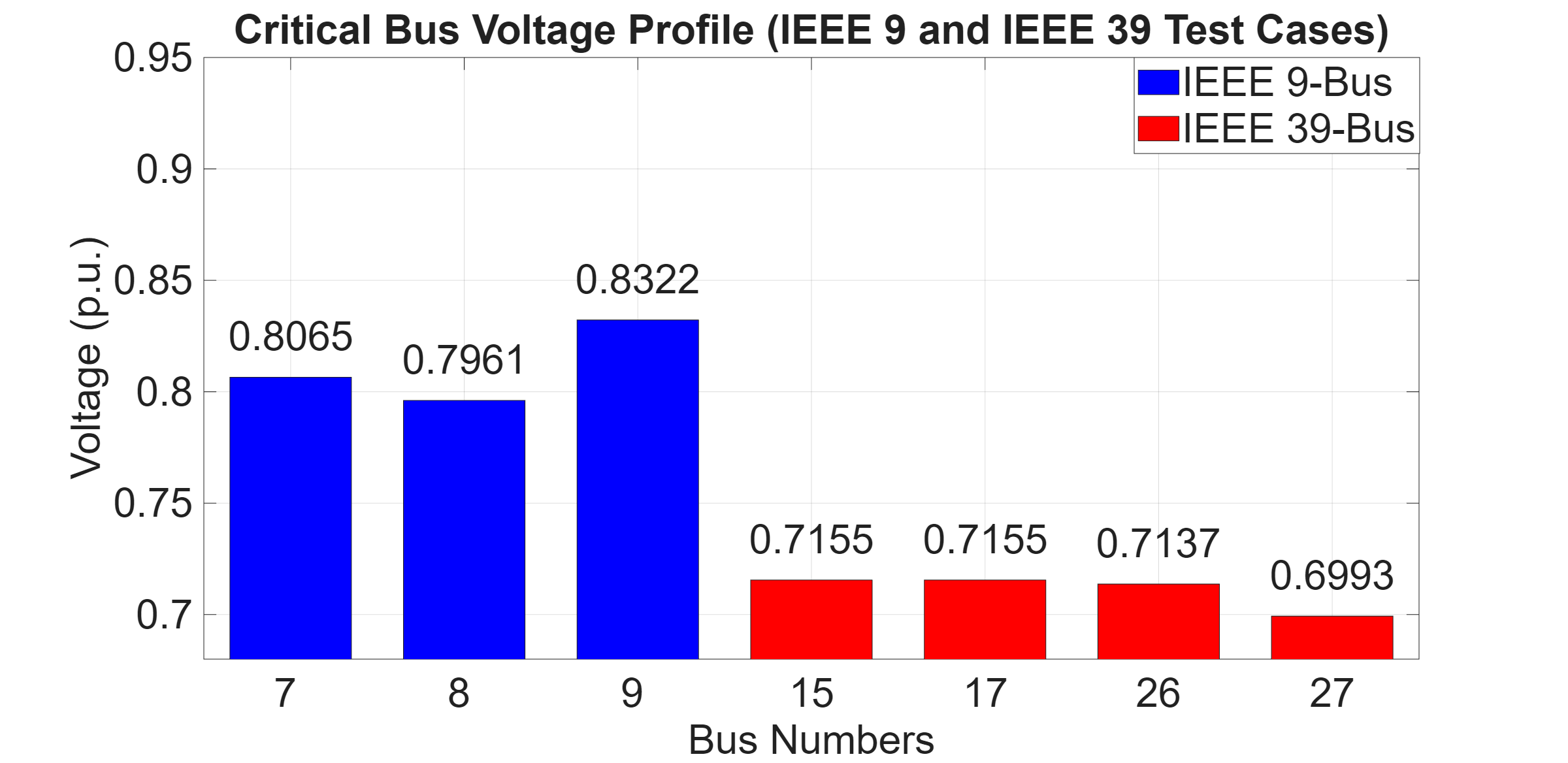}}
\caption{Voltage Profile of IEEE-9 Bus System and IEEE-39 Bus System (Critical Power Point).}
\label{fig}
\end{figure}

A stronger grid impact occurs with multiple EV load model integrations. Fig. 4 presents comparative voltage stability curves with tags marking critical power point values, voltage collapse, and grid integration effects of different EV models on critical buses for the IEEE 9-bus and IEEE 39-bus systems.

\begin{figure}[htbp]
\centerline{\includegraphics[width=3.6in,height=2in]{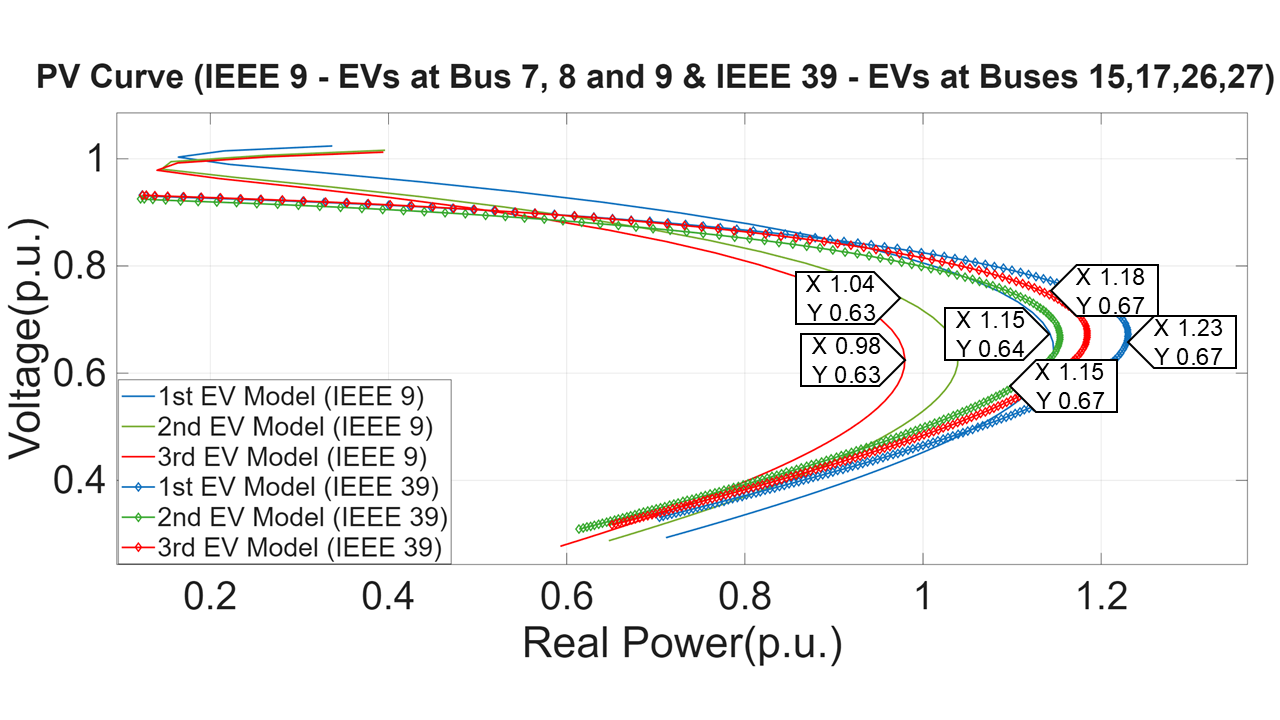}}
\caption{PV Curves for IEEE 9 Bus System (Critical Bus 6) and IEEE 39 Bus System (Critical Bus 12).}
\label{fig}
\end{figure}

The first EV model at critical buses shows better stability performance, with noticeable differences in loadability compared to the second model. Considering the charging limit for reactive power output and SoC control in the first model accounts for these differences in the stability curves. The third model shows moderate load variation with a less steep PV curve than the other models. Plotting PV curves of different transmission networks together helps to visualize the variation of grid impact from EV models during grid congestion.

\subsection{Defined solutions of voltage stability}
Based on EV load placements at critical buses, it is observed that the stability indices of line 7–8 for the third EV model are comparatively lower, identifying it as the weak transmission line for all EV model cases leading to grid instability. However, from the bar chart of stability indices in Fig. 5, the first EV model shows the highest critical reactive power values, making it distinct from the other models.

\begin{figure}[htbp]
\centerline{\includegraphics[width=3.3in,height=1.5in]{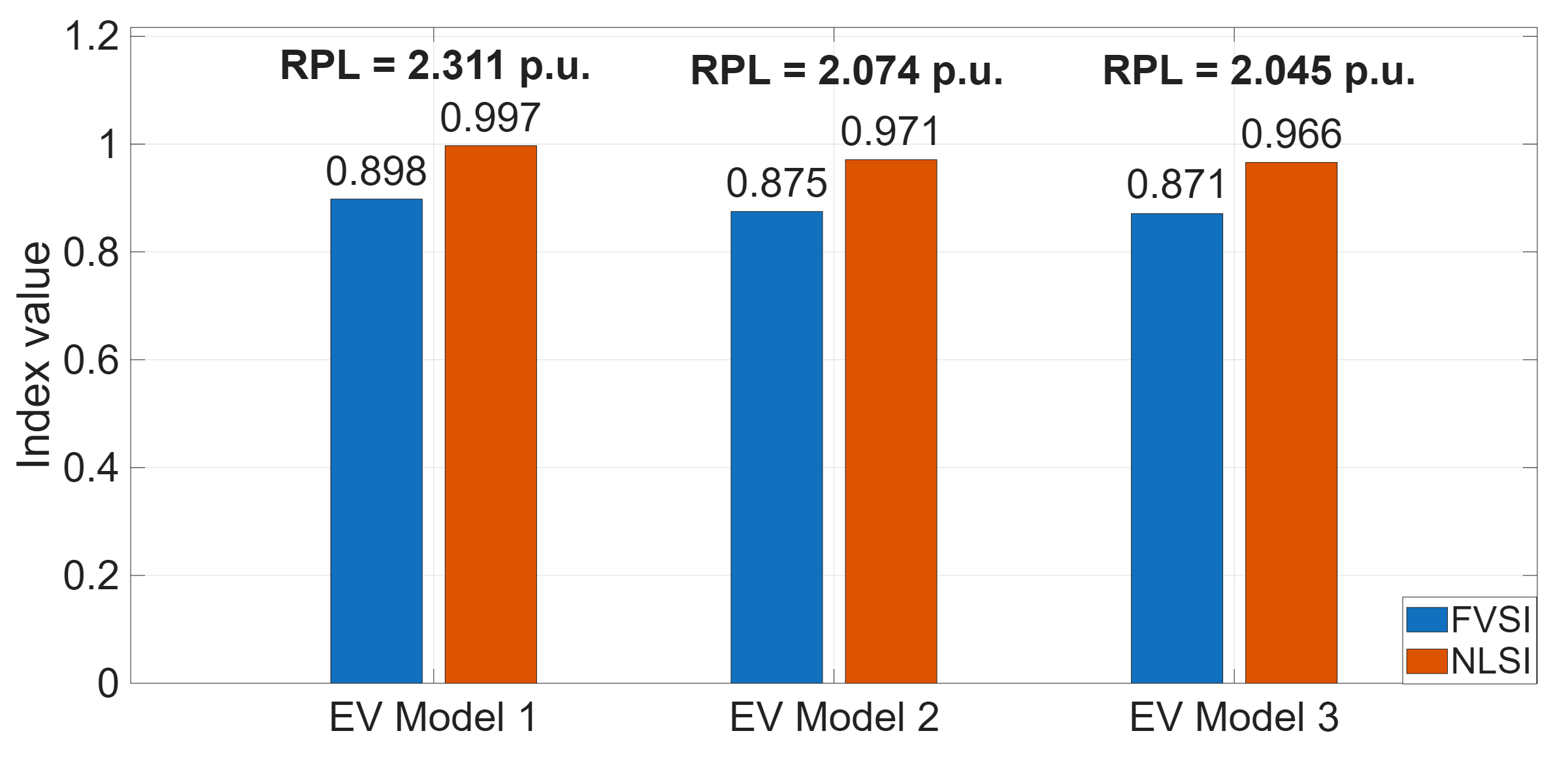}}
\caption{Stability Index of Weak Transmission Line 7-8 for EV models on Buses 7, 8 and 9 (IEEE 9 Bus).}
\label{fig}
\end{figure}

From the line stability index values in Fig. 6, line 14–15 is identified as the weak transmission line, indicating reactive power loading for different cases of all EV models integrated into the network. However, the third EV model shows the highest index values, tending toward instability at lower reactive power loadability. This highlights the importance of charging infrastructure and PQ control approaches used in the first and second EV models. Moreover, the first model performs better at higher critical reactive power compared to the second model.

\begin{figure}[htbp]
\centerline{\includegraphics[width=3.3in,height=1.5in]{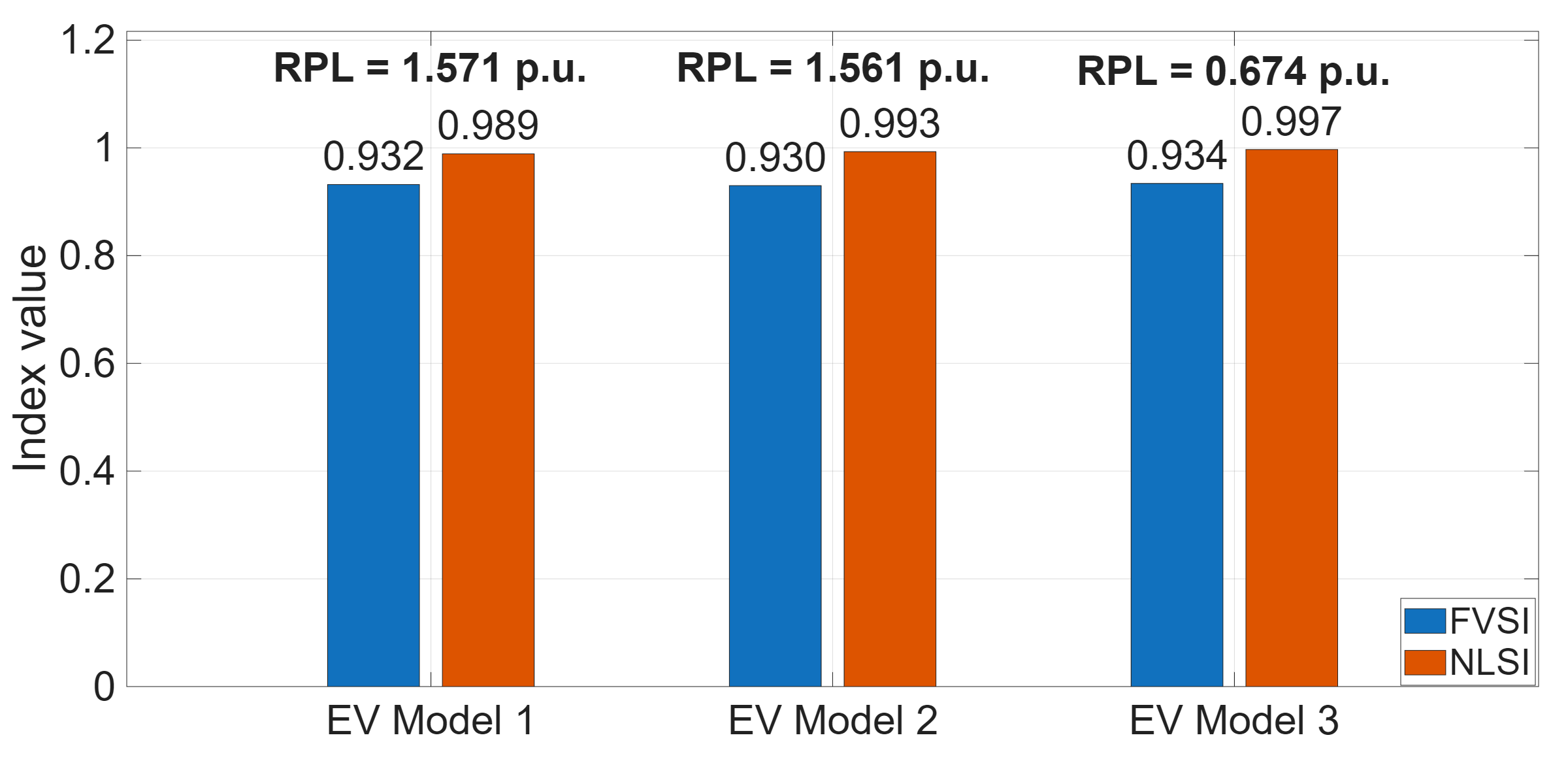}}
\caption{Stability Index of Weak Transmission Line 14-15 for EV models on Buses 15, 17, 26 and 27 (IEEE 39 Bus).}
\label{fig}
\end{figure}

\subsection{Frequency stability results with EV load grid integrations}

In these study scenarios, the IEEE 9-bus generator frequency profile
with proposed different EV load model integrations is analyzed. By applying MB PSS, the frequency profile for each EV load model is tightly synchronized, and the overall damping has been improved after the switching event of EV station loads occurred at 16th second to determine point of disturbance. Highest frequency oscillations for the second EV load model have been observed after the disturbance: Frequency Nadir has been estimated at 59.12 Hz of Generator 3 (PV Bus). Frequency profile for each EV model is represented in Fig. 7. 

Electromechanical oscillations would decay rapidly and function better for the case of the first EV load model, which enhances overall frequency stability among generators. Table – 2 and Table – 3 demonstrate the Frequency Nadir and ROCOF obtained from the frequency regulation of each synchronous generator. The reason behind using heat map format of these tables is to indicate highest and lowest stability factors, especially useful for Frequency Nadir values to comply with nominal grid frequency ranges. Overall, it is observed that the first EV load model has better nadir values compared to the other EV models, even though the better ROCOF values are estimated for the case of the third EV load model. 

\begin{figure}[htbp]
\centerline{\includegraphics[width=3.6in,height=1.9in]{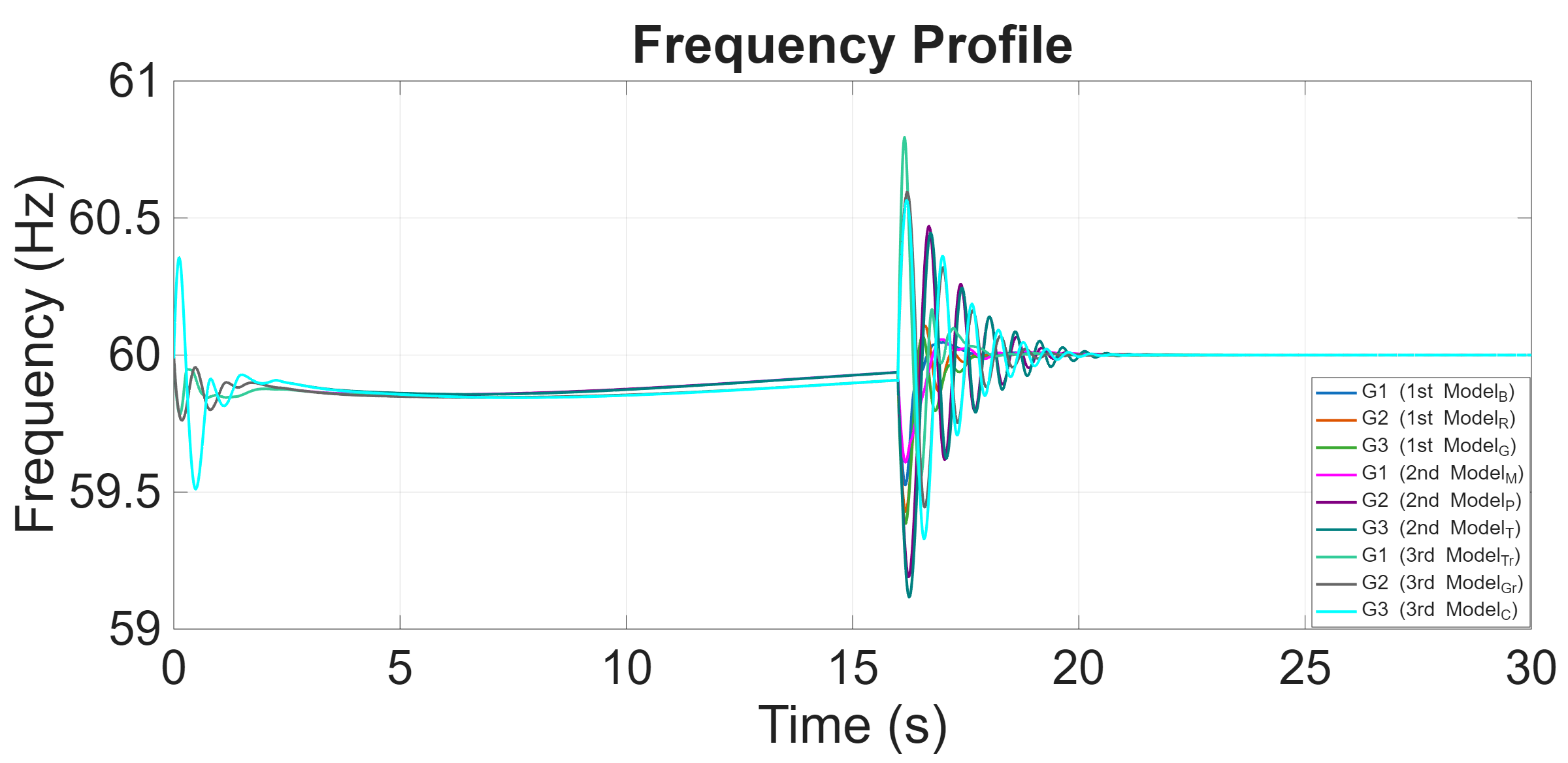}}
\caption{Frequency Response with EV Model Integrations (Bus 7, 8, 9) with MB PSS.}
\label{fig}
\end{figure}

\begin{table}[htbp]
  \caption{Frequency Nadir (Hz) - MB PSS for EV Models at Bus 7, 8, 9 (Deterministic Analysis).}
  \label{tab:freq-nadir-s4}
  \centering
  \setlength{\tabcolsep}{4pt}
  \renewcommand{\arraystretch}{1.2}
  \resizebox{\columnwidth}{!}{%
  \begin{tabular}{lccc}
    \toprule
    \textbf{Number of Generators} & \textbf{First EV model} & \textbf{Second EV model} & \textbf{Third EV model} \\
    \midrule
    \textbf{Generator 1} &
    \cellcolor{EVgreenA} 59.53 &
    \cellcolor{EVgreenB} 59.61 &
    \cellcolor{EVgreenC} 59.54 \\
    \textbf{Generator 2} &
    \cellcolor{EVyellowA} 59.43 &
    \cellcolor{EVsalmon} 59.19 &
    \cellcolor{EVyellowB} 59.44 \\
    \textbf{Generator 3} &
    \cellcolor{EVorangeA} 59.38 &
    \cellcolor{EVredA} 59.12 &
    \cellcolor{EVorangeB} 59.33 \\
    \bottomrule
  \end{tabular}%
  }
\end{table}

\begin{table}[htbp]
  \caption{ROCOF (Hz/s) - MB PSS for EV Models at Bus 7, 8, 9 (Deterministic Analysis).}
  \label{tab:freq-nadir-s4}
  \centering
  \setlength{\tabcolsep}{4pt}
  \renewcommand{\arraystretch}{1.2}
  \resizebox{\columnwidth}{!}{%
  \begin{tabular}{lccc}
    \toprule
    \textbf{Number of Generators} & \textbf{First EV model} & \textbf{Second EV model} & \textbf{Third EV model} \\
    \midrule
    \textbf{Generator 1} &
    \cellcolor{EVyellow} $-0.13$ &
    \cellcolor{EVyellow} $-0.13$ &
    \cellcolor{EVgreen} $-0.09$ \\
    \textbf{Generator 2} &
    \cellcolor{EVsalmonA} $-0.20$ &
    \cellcolor{EVsalmonB} $-0.21$ &
    \cellcolor{EVygreen} $-0.12$ \\
    \textbf{Generator 3} &
    \cellcolor{EVred} $-0.22$ &
    \cellcolor{EVred} $-0.22$ &
    \cellcolor{EVgreen} $-0.09$ \\
    \bottomrule
  \end{tabular}%
  }
\end{table}

From the stability parameters, it is observed that frequency response of first EV load model shows better frequency resilience with overall improved steady-state with respect to other EV load models. The overall settling time for each EV model are demonstrated briefly in Table - 4. It is also evident that the location and type of EV models for grid integration affect the overall frequency stability and grid oscillations.

\begin{table}[t]
\centering
\caption{Settling Time (s) for EV Models at Bus 7, 8, 9.}
\label{tab:settling_time_s2}
\renewcommand{\arraystretch}{1.25}
\setlength{\tabcolsep}{6pt}
\setlength{\arrayrulewidth}{0.6pt}
\resizebox{\columnwidth}{!}{%
\begin{tabular}{|c|c|c|c|}
\hline
\textbf{Number of Generators} & \textbf{First EV Model} & \textbf{Second EV Model} & \textbf{Third EV Model} \\
\hline
Generator 1 & 19.83 & 20.47 & 19.80 \\
\hline
Generator 2 & 19.66 & 20.30 & 21.83 \\
\hline
Generator 3 & 18.04 & 20.69 & 22.99 \\
\hline
\end{tabular}%
}
\end{table}

\section{Conclusion}
This research paper summarizes various EV load model integration for transmission-level weak-grid voltage stability analysis. From the case studies, the first EV model—with coordinated charging control and reactive power support—shows a larger critical power point and improved voltage profile. Stability indices identify the most vulnerable line and confirm that EV load placement is well organized within stability limits. The results recommend substantial EV load integration in weak grids without violating constraints, remaining feasible with the base load. The FVSI anticipates bus voltage sag, while the NLSI maintains indices near the threshold considering real and reactive power flow. For the IEEE 9-bus network, frequency stability is highly sensitive to PSS use and EV load placement; with EV integration, PSS improves control as oscillations decay faster, keeping the slack-bus generator’s frequency nadir near nominal with the lowest ROCOF value. Overall, the MB-PSS enhances frequency stability and post-disturbance settling. Future work can extend these findings to converter-based transient stability, protection systems, and FACTS devices, as well as the IEEE 39-bus network.

\section*{Acknowledgement}
The authors express their sincere gratitude to the support provided by Electrical Engineering Department at King Fahd University of Petroleum and Minerals (KFUPM) in conducting this research work. The authors extend their sincere appreciation to the Interdisciplinary Research Center for Sustainable Energy Systems (IRC-SES) at KFUPM.


\begin{thebibliography}{00}

\bibitem{b1} A. Tavakoli, S. Saha, M. T. Arif, M. E. Haque, N. Mendis, and A. M. T. Oo, “Impacts of grid integration of solar PV and electric vehicle on grid stability, power quality and energy economics: A review,” IET Energy Systems Integration, vol. 2, no. 3, pp. 215–225, Sep. 2020, doi: 10.1049/iet-esi.2019.0047.
\bibitem{b2} P. Kundur et al., “Definition and classification of power system stability,” IEEE Transactions on Power Systems, vol. 19, no. 3, pp. 1387–1401, Aug. 2004, doi: 10.1109/TPWRS.2004.825981.
\bibitem{b3} Enerdynamics, ``How Electric Operators Maintain Acceptable Voltage,'' accessed Mar.~13, 2025. [Online]. Available:
\url{https://www.enerdynamics.com/Energy-Currents_Blog/How-Electric-Operators-Maintain-Acceptable-Voltage.aspx}
\bibitem{b4} “Power Factor: Determining how Much Electricity Your Power System Consumes - Technical Articles.” Accessed: Mar. 13, 2025. [Online]. Available: https://eepower.com/technical-articles/power-factor-determining-how-much-electricity-your-power-system-consumes/
\bibitem{b5} M. Alzubaidi, K. N. Hasan, and L. Meegahapola, “Probabilistic steady-state and short-term voltage stability assessment considering correlated system uncertainties,” Electric Power Systems Research, vol. 228, Mar. 2024, doi: 10.1016/j.epsr.2023.110008.
\bibitem{b6} A. M. Refat, W. M. Hamanah, A. S. Menesy, M. A. Abido, and Md. Shafiullah, “Voltage Stability Assessment of Renewable Microgrids Using Deep Learning,” in 2025 IEEE 15th International Conference on Power Electronics and Drive Systems (PEDS), IEEE, Jul. 2025, pp. 1–5. doi: 10.1109/PEDS63958.2025.11144830.
\bibitem{b7} K. Momoh, S. A. Zulkifli, P. Korba, F. R. S. Sevilla, A. N. Afandi, and A. Velazquez-Ibañez, “State-of-the-Art Grid Stability Improvement Techniques for Electric Vehicle Fast-Charging Stations for Future Outlooks,” May 01, 2023, MDPI. doi: 10.3390/en16093956.
\bibitem{b8} “Frequency Response Initiative Report The Reliability Role of Frequency Response,” 2012. [Online]. Available: www.nerc.com
\bibitem{b9} D. Vásquez-Cardona, S. D. Saldarriaga-Zuluaga, S. Bustamante-Mesa, J. M. López-Lezama, and N. Muñoz-Galeano, “Impacts of Electric Vehicle Penetration on the Frequency Stability of Curaçao’s Power Network,” World Electric Vehicle Journal, vol. 16, no. 5, May 2025, doi: 10.3390/wevj16050264.
\bibitem{b10} K. Rajashekaraiah, C. Iurlaro, S. Bruno, and G. De Carne, “Modelling of 3-Phase p-q Theory-Based Dynamic Load for Real-Time Simulation,” IEEE Open Access Journal of Power and Energy, vol. 10, pp. 654–664, 2023, doi: 10.1109/OAJPE.2023.3340299.
\bibitem{b11} H. Farhaj Khan, A. Hanif, and N. Anwar, “Rotor Angle and Voltage Stability Analysis with Fault Location Identification on the IEEE 9 Bus System,” 2020. [Online]. Available: www.etasr.com
\bibitem{b12} H. S. Salama and I. Vokony, “Voltage stability indices–A comparison and a review,” Computers and Electrical Engineering, vol. 98, Mar. 2022, doi: 10.1016/j.compeleceng.2022.107743.
\bibitem{b13} F. M. Aboshady, I. Pisica, A. F. Zobaa, G. A. Taylor, O. Ceylan, and A. Ozdemir, “Reactive Power Control of PV Inverters in Active Distribution Grids with High PV Penetration,” IEEE Access, vol. 11, pp. 81477–81496, 2023, doi: 10.1109/ACCESS.2023.3299351.
\bibitem{b14} “Systems-Level Microgrid Simulation from Simple One-Line Diagram - File Exchange - MATLAB Central.” Accessed: Nov. 10, 2025. [Online]. Available: https://www.mathworks.com/matlabcentral/fileexchange/67060-systems-level-microgrid-simulation-from-simple-one-line-diagram
\bibitem{b15} \emph{Academia.edu}, "Small-Signal Stability of an Islanded Microgrid." Accessed: Nov. 6, 2025. [Online]. Available: \url{https://www.academia.edu/98711088/Small_Signal_Stability_Of_An_Islanded_microgrid}
\bibitem{b16} C. H. Dharmakeerthi, N. Mithulananthan, and T. K. Saha, “Impact of electric vehicle fast charging on power system voltage stability,” International Journal of Electrical Power and Energy Systems, vol. 57, pp. 241–249, May 2014, doi: 10.1016/j.ijepes.2013.12.005.
\bibitem{b17} “Automatic Generation Control by Wind generation resource co-ordinated with synchronous generator - File Exchange - MATLAB Central.” Accessed: Sep. 23, 2025. [Online]. Available: https://www.mathworks.com/matlabcentral/fileexchange/53764-automatic-generation-control-by-wind-generation-resource-co-ordinated-with-synchronous-generator
\bibitem{b18} \emph{MathWorks}, "Multiband Power System Stabilizer — Implement multiband power system stabilizer — Simulink." Accessed: Sep. 23, 2025. [Online]. Available: \url{https://www.mathworks.com/help/sps/powersys/ref/multibandpowersystemstabilizer.html}
\bibitem{b19} M. El-Sadek, G. Shabib, Y. Mobarak, and M. El-Ahmar, “COMBINED CONTROLS OF STATCOM DEVICE AND MULTI-BAND POWER SYSTEM STABILIZER (MB-PSS) IN POWER SYSTEM,” JES. Journal of Engineering Sciences, vol. 37, pp. 115–124, Jan. 2009, doi: 10.21608/jesaun.2009.121027.

\end{thebibliography}
\end{document}